\documentclass[a4paper]{jpconf}
\usepackage{amsmath,amssymb,amsfonts}
\usepackage{dbnsymb}
\usepackage{graphicx}

\def\ben{\begin{equation}}
\def\een{\end{equation}}
\def\bea{\begin{eqnarray}}
\def\eea{\end{eqnarray}}
\def\half{\frac{1}{2}}
\def\R{{{\mathbb R}}}

\begin{document}
\title{Classical GR as a topological theory with linear constraints\footnote{Based on a talk given at the Spanish Relativity Meeting (ERE2010), Granada, on 6 September 2010 \cite{talkpdf}.}}

\author{Steffen Gielen}

\address{DAMTP, University of Cambridge, Centre for Mathematical Sciences, Wilberforce Road, Cambridge CB3 0WA, U.K.}
\address{Max Planck Institute for Gravitational Physics (Albert Einstein Institute), Am M\"uhlenberg 1, D-14476 Golm, Germany}

\ead{sg452@cam.ac.uk, gielen@aei.mpg.de}

\begin{abstract}
We investigate a formulation of continuum 4d gravity in terms of a constrained topological (BF) theory, in the spirit of the Plebanski formulation, but involving only linear constraints, of the type used recently in the spin foam approach to quantum gravity.
We identify both the continuum version of the linear simplicity constraints used in the quantum discrete context and a linear version of the quadratic volume constraints that are necessary to complete the reduction from the topological theory to gravity. We illustrate and discuss also the discrete counterpart of the same continuum linear constraints.
Moreover, we show under which additional conditions the discrete volume constraints follow from the simplicity constraints, thus playing the role of secondary constraints.
Our analysis clarifies how the discrete constructions of spin foam models are related to a continuum theory with an action principle that is equivalent to general relativity.
\end{abstract}

\section{Motivation}
As is well known, general relativity in three dimensions is topological, as there are no local degrees of freedom. Its first order formulation in terms of a connection $\omega^{AB}$ and a triad $E^A$ is given by
\ben
S=\frac{1}{16\pi G}\int_{\Sigma} \epsilon_{ABC} E^A\wedge R^{BC}[\omega]\,.
\een
The quantisation of such a theory is rather well understood and different possible quantisation procedures are known. This is of course in stark contrast to general relativity in four dimensions, whose quantisation is to a large extent still an open problem. One may try to exploit what is known in 3D to write general relativity in 4D as a {\it constrained} topological theory with action
\ben
S=\int_{\Sigma} B^{AB}\wedge R_{AB}[\omega] + \lambda^{\alpha}C_{\alpha}[B]\,;
\een
one then needs to add constraints (the ``simplicity constraints") which enforce $B^{AB}=\frac{1}{16\pi G}{\epsilon^{AB}}_{CD}E^C\wedge E^D$ (for some set of 1-forms $E^A$) to recover general relativity. The main obstacle in quantisation is then the identification and implementation of such constraints, which has indeed been a topic of much recent debate in the discrete setting of spin foam models \cite{newmodels}.

If, for example as the starting point for loop quantum gravity, one wants to add a Holst term \cite{holst} to the first order connection form for general relativity, one instead needs to impose $\Sigma^{AB}=\frac{1}{8\pi\gamma G}E^A\wedge E^B$ on the two-form field $\Sigma^{AB}$, defined in terms of $B^{AB}$ by 
\ben
\Sigma^{AB}=\frac{1}{1\pm\gamma^2}\left(B^{AB}-\frac{\gamma}{2}{\epsilon^{AB}}_{CD}B^{CD}\right)
\een
(note that this is just a linear redefinition). We will focus on this case in the following and try to determine an appropriate set of constraints for $\Sigma^{AB}$.

In the traditional (Plebanski) formulation \cite{plebanski}, one uses {\it quadratic constraints} for $\Sigma^{AB}$,
\ben
\epsilon_{ABCD}\Sigma^{AB}_{ab}\Sigma^{CD}_{cd}=V\epsilon_{abcd},
\een
which have two {\it separate} sectors of solutions\footnote{This is under the additional assumption that $V\neq 0$! $V=0$ configurations are not geometric at all, leading to additional potential difficulties in quantisation \cite{reisen}.}:
\ben
\mbox{either }\Sigma^{AB}=\pm e^A\wedge e^B\quad\mbox{or }\Sigma^{AB}=\pm\half{\epsilon^{AB}}_{CD}e^C\wedge e^D
\een
for some set of 1-forms $e^A$. Classically, one can consistently remain within the ``GR" sector \cite{reisen}; quantum mechanically, the situation is less clear.

The discrete analogue of this construction has been used in spin foam models of quantum gravity: One introduces a triangulation of spacetime, integrates $\Sigma^{AB}$ over triangles
\ben
\Sigma^{AB}_{ab}(x)\quad\Rightarrow\quad\Sigma^{AB}_{\triangle}\equiv\int_{\triangle}\Sigma^{AB}\in\frak{so}(4)\simeq\Lambda^2 \R^4,
\een
and imposes the constraint $\epsilon_{ABCD}\Sigma^{AB}_{\triangle}\Sigma^{CD}_{\triangle'}=0$ if $\triangle=\triangle'$ or $\triangle$ and $\triangle'$ share an edge. It can be shown \cite{livinespeziale} that the remaining ``volume" constraints can be replaced by the ``closure constraint"
\ben
\sum_{\triangle\subset\tetrahedron}\Sigma^{AB}_{\triangle}=0.
\label{closure}
\een
The quantum-mechanical implementation of these constraints led to the Barrett-Crane model \cite{bc}. The recently proposed new spin foam models \cite{newmodels}, on the other hand, rely on the replacement of quadratic constraints on $\Sigma^{AB}_{\triangle}$ by {\em linear} constraints:
\ben
n_A(\tetrahedron)\Sigma^{AB}(\triangle)=0\quad\forall\;\triangle\subset\tetrahedron,
\label{linearconst}
\een
where $n_A(\tetrahedron)$ is normal to the tetrahedron $\tetrahedron$. These are stronger than the quadratic constraints; they restrict $\Sigma^{AB}$ to the discrete analogue of $\Sigma^{AB}=\pm E^A\wedge E^B$. Here we present an extension of this construction, which was done in the discrete setting, to the {\it classical continuum theory}: We give a set of constraints whose solution implies that $\Sigma^{AB}=\pm E^A\wedge E^B$ for some $E^A$ (ignoring the factor $8\pi\gamma G$ in the following). We then discretise these constraints and identify some which are sufficient at the discrete level. For details and proofs of our claims, see \cite{paper}.

\section{The construction}

\subsection{Continuum construction}
Clearly, to find a continuum analogue of (\ref{linearconst}) that is derivable from an action, one has to introduce dynamical fields that play the analogue of the normal vectors $n^A$ in (\ref{linearconst}). We introduce a basis of 1-forms $e^A_a$ (i.e. assuming $\det (e^A_a)\neq 0$), which is dual to a basis of 3-forms by
\ben
n_{Adef}\equiv\epsilon_{ADEF}e^D_d e^E_e e^F_f,\quad e^c_A \sim \epsilon^{cdef}n_{Adef}.
\een
{\bf Claim 1.} For a basis of 3-forms $n_A$, the general solution to
\ben
n_{Adef}\Sigma^{AB}_{ab}=0\quad\forall\{a,b\}\subset\{d,e,f\}
\label{con}
\een
is 
\ben
\Sigma^{AB}_{ab}=G_{ab}e^{[A}_a e^{B]}_b,
\een
where $e^A$ is defined in terms of $n_A$ as above. Note that $G_{ab}=G_{ab}(x)$ are functions on spacetime. One could try a linear redefinition $e^A_a=\lambda_a E_a^A$ to identify this general solution
\ben
\Sigma^{AB}_{ab}=G_{ab}e^{[A}_a e^{B]}_b
\een
with $\Sigma^{AB}=\pm E^A\wedge E^B$, but this is not possible in general; one needs additional conditions. 

In \cite{paper}, we show that imposing the additional three constraints
\ben
\sum_b\sum_{\{a,f\}\not\in\{b,e\}}n_{Abef}\Sigma^{AB}_{ab}=0,\quad e\in\{0,1,2\}\;\mbox{fixed,}
\label{addcon}
\een
implies that $G_{ab}(x)$ is the same function $c(x)$ for all $a$, $b$, which can be absorbed by a pointwise rescaling $E^A=\sqrt{|c|}e^A$ to complete the identification $\Sigma^{AB}=\pm E^A\wedge E^B$. Note that not the field $e^A$ we have introduced to define the constraints (\ref{con}), but the rescaled $E^A$ will play the role of an orthonormal tetrad determining physical lengths and angles. In particular, a completely degenerate triad $E^A=0$ is possible since $c(x)=0$ at some points is not excluded.

\subsection{Discrete construction}
The translation of the constraints into discretised variables is straightforward. Again one introduces a triangulation of spacetime and integrates $\Sigma^{AB}$ over triangles and $n_A$ over tetrahedra:
\ben
\Sigma^{AB}_{ab},\;n_{Adef}\;\Rightarrow\;\Sigma^{AB}(\triangle),\;n_A(\tetrahedron).
\een
Then the discrete analogue of the set of constraints (\ref{con}) is (essentially by construction) the set of linear constraints used in the EPR(L)/FK models, $n_A(\tetrahedron)\Sigma^{AB}(\triangle)=0\quad\forall\;\triangle\subset\tetrahedron$. Perhaps more interestingly, the discrete version of the remaining constraints (\ref{addcon}) would take the form
\ben
\sum_{\{i,j\}\not\ni {\bf A}}n_A(\tetrahedron_{{\bf i}})\Sigma^{AB}(\triangle_{{\bf Aj}})=0,
\label{volanalogue}
\een
where we label the tetrahedra in a 4-simplex by ${\bf A, B, C, D, E}$; there are five of these constraints per simplex (where ${\bf A}$ is replaced by ${\bf B, C, D, E}$ respectively). These constraints play the role of the ``volume" constraints which could be, in the case of quadratic constraints, replaced by a closure constraint relating different triangles in a tetrahedron. The situation here is completely analogous:
\\
\\{\bf Result:} The constraints (\ref{volanalogue}) follow from the EPR(L)/FK linear constraints, the closure constraint on $\Sigma^{AB}(\triangle)$, plus an analogous ``4D closure constraint"
\ben
n_A(\tetrahedron_{{\bf A}})+n_A(\tetrahedron_{{\bf B}})+n_A(\tetrahedron_{{\bf C}})+n_A(\tetrahedron_{{\bf D}})+n_A(\tetrahedron_{{\bf E}})=0,
\label{4dconst}
\een
which can be given a clear geometric motivation, just as closure on $\Sigma^{AB}$: While the closure constraint (\ref{closure}) can be interpreted as the requirement for the triangles specified by $\Sigma^{AB}_{\triangle}$ to close up to form a tetrahedron, or alternatively as the discrete version of the field equation $\nabla_{[a}\Sigma^{AB}_{bc]}=0$ integrated over an infinitesimal 3-ball, the four-dimensional constraint (\ref{4dconst}) can be interpreted as demanding that the five tetrahedra specified by $n^A(\tetrahedron)$ close up to form a 4-simplex. The derivation of (\ref{4dconst}) from a continuum field equation is more subtle since one has to assume both $\nabla_{[a}\Sigma^{AB}_{bc]}=0$ and the identification of $\Sigma^{AB}$ with $e^A$ to obtain the equation $\nabla_{[a}n^{A}_{bcd]}=0$.

\section{Summary}
We have established a formulation of classical general relativity as a topological BF theory plus linear constraints which relied on the introduction of a basis of 3-forms at each point in addition to the two-form field $\Sigma^{AB}$ and the connection $\omega^{AB}$. This new continuum action is more closely related to current spin foam models, such as the EPR(L) and FK models, as discretising it leads to variables $\Sigma^{AB}(\triangle),\;n_A(\tetrahedron)$ which appear in these spin foam models, and are subject to the constraints used in these models, plus closure constraints on both $\Sigma^{AB}$ and $n_A$. The latter is a new constraint which has been suggested before, but not received much attention in the literature so far. It suggests a new formulation for spin foam models in which the normals $n_A$ carry information about the 3-volume and thus are given a fully geometric role, other than in the current models where one can set them to a fixed value. Other than searching for a modification of spin foam models in this direction, one should perform a detailed canonical analysis of the constraints used in our formulation. There are also indications for a close relation of our constructions both to current research in group field theories, as well as to the ``edge simplicity" constraints introduced in \cite{biancajimmy}. We leave all this to future work.

\ack
This work was supported by EPSRC and Trinity College, Cambridge, partly through a Rouse Ball Travelling Studentship in Mathematics.

\end{document}